\documentstyle[preprint,aps]{revtex}
\begin{document}
\preprint{TP-MUBR 93-11/1}
\draft
\title{Finite temperature dilaton gravity}
\author{Zden\v{e}k Kopeck\'{y}\cite{email}}
\address{Department of Theoretical Physics and Astrophysics,
Masaryk University,\\ Kotl\'a\v{r}sk\'{a} 2, 611 37 Brno, Czech Republic}
%\date{\today}
\maketitle
\begin{abstract}
The dilaton free energy density in external static gravitational field
is found. We use the real time formulation of the finite temperature field
theory and the free energy density is computed to the first order of the
string parameter $\alpha '$. We obtain the  thermal corrections to
the $\alpha'$ modified Einstein gravity action.
\end{abstract}
% body of paper here
\narrowtext

\section{Introduction}

The high-temperature properties of quantum gravity are of some interest
for their potential cosmological applications. The modifications of the
Einstein action have been done in the frame both the finite
temperature theory and the string theory.

The finite temperature theory yields the thermal-quantum corrections
(by gravitons and others particles) to the effective gravity action
\cite{ThermGrav0,ThermGrav1,ThermGrav2,ThermGrav3}.

In standard treatments of string theory, it is shown that a consistent
string theory can be formulated from some class of conformally
invariant models.
The  $\alpha'$ corrections to the Einstein action are known for strings
moving in  background fields \cite{StrBack,StrBackCMP}.

Some aspects of the  $\alpha'$ modified  black hole hermodynamics
have been recently investigated in  Ref. \cite{HarmLeb}.

In this paper, we derive the dilaton free energy density in curved background
which is considered static. The free energy density with opposite sign
is contribution to the total gravity Lagrangian.
The action for the dilaton  field is taken from string theory
to order $O(\alpha')$.

\section{Dilaton action}

Corrections to the external metric field action due to stringy effect were
carried out to $O(\alpha')$ in Ref. \cite{StrBackCMP}. The action has form for
bosonic string case
\begin{eqnarray}
S_0=\frac{1}{16\pi}\int d^{D}x\sqrt{-g}\,e^{-2\phi}(R-4(\nabla\phi)^{2}
+\frac{\alpha'}{4}R_{klmn}R^{klmn})\; ,
\label{eq:SB}
\end{eqnarray}
where  D=26 is the dimension corresponds to the critical bosonic
string theory, $g_{ik}(x)$ and $\phi(x)$ are the metric tensor field
and the dilaton field, respectively. We consider situation without
the antisymmetrical background field \cite{StrBack,StrBackCMP} in
the action Eq.\ (\ref{eq:SB}).
The conformal transformation,
\begin{eqnarray}
g_{ij}\rightarrow \exp(\frac{4}{D-2}\phi)\,g_{ij}\; ,
\label{eq:DTrans}
\end{eqnarray}
together with some field redefinitions, changes the action (\ref{eq:SB}) to
\cite{StrBackCMP}
\begin{eqnarray}
S_0=\frac{1}{16\pi}\int d^{D}x\sqrt{-g}\,(R+\frac{4}{D-2}(\nabla\phi)^{2}
+\frac{\alpha'}{4}e^{\frac{-4\phi}{D-2}}R_{klmn}R^{klmn})\; .
\label{eq:DATr}
\end{eqnarray}
We will assume that all but four space-time dimensions are compactified out
in the action (\ref{eq:SB}).
In the rest of the article, we consider the four dimensional manifold with
signature $(+,-,-,-)$. Related space-time indexes  and  space indexes
have ranges
$\mu,\nu,...=0...3$
and   $i,j,...=1...3$, respectively.
The dilaton part of the action Eq.\ (\ref{eq:DATr})  we identify with
\begin{eqnarray}
S[\phi]=\int d^{4}x\sqrt{-g}\,(\frac{1}{8\pi\,G}(\nabla\phi)^{2}+
\frac{\alpha'}{64\pi\,G}e^{-2\phi}
R_{\mu\nu\tau\sigma}R^{\mu\nu\tau\sigma})\; .
\label{eq:DilAc4}
\end{eqnarray}
We have written out explicitly the gravitational constant in
Eq.\ (\ref{eq:DilAc4}).

\section{Dilaton free energy to order $O(\alpha')$}

In general, statistical mechanics in static space-time  can  be developed
by  constructing the following partition function \cite{DowKen,ThermI}
\begin{eqnarray}
Z[\beta]=Tr \exp(-\beta\,\hat{H})\; .
\end{eqnarray}
The Hamiltonian $\hat{H}$ is connected with
$\hat{T}_{0}^{0}$ component of the energy momentum tensor
\begin{eqnarray}
\hat{H}=\int\limits_{x_{0}=const.} d^{3}x\; \sqrt{-g}\,\hat{T}_{0}^{0}\; .
\end{eqnarray}
The parameter $\beta$ corresponds to the temperature $\beta^{\nu}$ vector
\cite{ThermI}
in  static coordinates
\begin{equation}
\beta^{\nu}=(\beta,{\bf 0})\; .
\end{equation}
The local Lorentz-rest-frame inverse (scalar) temperature $\beta_{R}$
then is
\begin{equation}
\beta_{R} = \sqrt{g_{00}}\;\beta \; .
\label{eq:betaR}
\end{equation}

Now, according to the above construction, we consider
dilaton field described by Eq.\ (\ref{eq:DilAc4}).
The general path integral formulation for the partition function in
the Real Time Finite Temperature
Theory (RTFT) is  \cite{RTFT,Evans}
\begin{eqnarray}
Z[\beta]&&=\int D\phi\, \exp(i\,S[\phi])\nonumber\\
&&=\exp(\, -\beta \int\limits_{x^{0}=const.} d^{3}x\sqrt{-g}\,F[g]\,) \; .
\label{eq:Z}
\end{eqnarray}
The action $S[\phi]$ is given by Eq.\ (\ref{eq:DilAc4}).
The free energy density $F[g]$  has been introduced in Eq.\ (\ref{eq:Z}).
The time integration $\int dx^{0}$ in the action $S[\phi]$ in
(\ref{eq:Z}) is going along curve in complex plane \cite{RTFT,Evans}.
The curve is for the RTFT
\begin{eqnarray}
C_{RTFT}=(-t,t) \cup (t,t-i\frac{\beta}{2})
                &&\cup (t-i\frac{\beta}{2},-t-i\frac{\beta}{2})\nonumber\\
                &&\cup (-t-i\frac{\beta}{2},-t-i\beta)
\label{eq:C}
\end{eqnarray}
and $ t\rightarrow \infty$.
The integration $\int D\phi$ in (\ref{eq:Z}) is over fields which satisfy the
periodicity condition
\begin{equation}
\phi(-t,{\bf x})=\phi(-t-i\beta,{\bf x})
\end{equation}
at the endpoints of the curve (\ref{eq:C}).

The interaction part of the action (\ref{eq:DilAc4}) we identify with
the second exponential term in Eq.\ (\ref{eq:DilAc4}).
The free energy density then has the first two terms of the expansion
to order O($\alpha'$)
\begin{eqnarray}
F[g]=F_{0}[g]-
\frac{\alpha'}{64\pi G}R_{\mu\nu\tau\sigma}R^{\mu\nu\tau\sigma}
(V(\beta,g)+1)\; ,
\label{eq:VF}
\end{eqnarray}
where
\begin{eqnarray}
V(\beta,g)&=&\sum_{n=1}^{\infty}(-2)^{2n}\frac{1}{n!}
\frac{1}{2^{n}}(i\Delta[\beta]_{11}(x,x))^{n}\nonumber\\
          &=&\exp(2i\Delta[\beta]_{11}(x,x))-1\; .
\label{eq:V}
\end{eqnarray}
The term $F_{0}[g]$ is contribution without the string corrections
\begin{eqnarray}
F_{0}[g]=-\int dm^2\frac{1}{2}\frac{i\Delta[\beta]_{11}(x,x)}{4\pi G}\; .
\label{eq:F0m}
\end{eqnarray}
The function $\Delta[\beta]_{11}(x,x)$ is the $(1,1)$ component of the RTFT
causal Green's function (\ref{eq:Delta}) calculated in the Appendix.
We have used the RTFT methods for the computation of the thermal vacuum
diagrams \cite{RTFT,Evans} to evaluate the term (\ref{eq:V}).
The term $V(\beta,g)$ is the sum of the closed connected vacuum graphs,
each graph has one vertex with fixed the thermal index to $1$
\cite{RTFT,Evans}
and the vertex is common for $n=1,2,3...$ number of the thermal propagator
function $\Delta[\beta]_{11}(x,x)$ loops.
Substitution Eqs.\ (\ref{eq:Delta}),(\ref{eq:DeltaL}) in Eq.\ (\ref{eq:F0m})
yields
\begin{eqnarray}
F_{0}[g]=-\frac{1}{32\pi^{2}}
\int\limits_{0}^{\infty} \frac{ds}{s^3} \sum_{k=0}^{\infty} a_{k}(x) s^{k}
\sum_{n=-\infty}^{\infty}
e^{\frac{-n^2\beta_{R}^{2}}{4s}}\; .
\label{eq:F0I}
\end{eqnarray}
We have taken limit $m \rightarrow 0$ in Eq.\ (\ref{eq:F0I}).
Using analytic continuation relation
\begin{equation}
\int\limits_{0}^{\infty} ds s^{z}e^{-as}=\frac{\Gamma(z+1)}{a^{z+1}}\; .
\end{equation}
and the $\zeta$ function regularization \cite{BirDev}
we transform Eq.\ (\ref{eq:F0I}) to
\begin{eqnarray}
F_{0}[g]
=&&-\frac{1}{(4\pi)^2}\sum_{k=0}^{\infty}a_{k}(x)\Gamma(2-k)\zeta(4-2k)
(\frac{2}{\beta_{R}})^{2(2-k)}\nonumber\\
&&-\lim_{\nu\rightarrow 1}\frac{1}{32\pi^{2}}\frac{a_{2}(x)}{\nu-1}
+\frac{1}{64\pi^2}\int\limits_{0}^{\infty}
ids\, \ln(\frac{is}{8\pi^{2}G})\,\frac{\partial^3}{\partial(is)^3}
\Lambda(x,is)\; .
\label{eq:F0}
\end{eqnarray}
The zero temperature term with the pole is canceled against the k=2 term
in Eq.\ (\ref{eq:F0}) because
\begin{equation}
\lim_{\nu\rightarrow 1}\Gamma(\nu-1)\zeta(0)=
\lim_{\nu\rightarrow 1} \frac{-1}{2(\nu-1)}\; .
\end{equation}
The $k>2$ terms in Eq.\ (\ref{eq:F0}) are finite because of the following
relation for the product of the $\zeta$ and $\Gamma$ function
\begin{equation}
\zeta(2z)\Gamma(z)=
4^{z}\pi^{2z}\,\frac{\Gamma(1-2z)}{\Gamma(1-z)}\,\zeta(1-2z)\; .
\label{eq:ZG}
\end{equation}
Final form of  Eq.\ (\ref{eq:F0}) using Eq.\ (\ref{eq:ZG}) will be
\begin{eqnarray}
F_{0}[g]
=&&\frac{-\pi^{2}}{90\beta_{R}^{4}}-a_{1}(x)\frac{1}{24\beta_{R}^{2}}
\nonumber\\
&&-\frac{1}{16\pi^{2}}\sum_{k=3}^{\infty}a_{k}(x)
\frac{(2k-4)!}{(k-2)!}\zeta(2k-3)(\frac{\beta_{R}}{4\pi})^{2(k-2)}\nonumber\\
&&+\frac{1}{64\pi^2}\int\limits_{0}^{\infty}
ids\, \ln(\frac{is}{8\pi^{2}G})\,\frac{\partial^3}{\partial(is)^3}
\Lambda(x,is)\; .
\label{eq:F0f}
\end{eqnarray}
The Green's function  $\Delta[\beta]_{11}(x,x)$ for $m \rightarrow 0$
can be written similarly
\begin{eqnarray}
\frac{i\Delta[\beta]_{11}(x,x)}{4\pi G}
=\frac{1}{12\beta_{R}^{2}}
+\frac{1}{8\pi^{2}}
\sum_{k=2}^{\infty}\,a_{k}(x)\,\frac{(2k-2)!}{(k-1)!}
\zeta(2k-1)\,(\frac{\beta_{R}}{4\pi})^{2(k-1)}\; .
\label{eq:Gf}
\end{eqnarray}
Now we can collect all the terms in Eq.\ (\ref{eq:VF}) and write final result
for the dilaton free energy density
\widetext
\begin{eqnarray}
F[g]
=&&\frac{-\pi^{2}}{90\beta_{R}^{4}}-a_{1}(x)\frac{1}{24\beta_{R}^{2}}
-\frac{1}{16\pi^{2}}\sum_{k=3}^{\infty}a_{k}(x)
\frac{(2k-4)!}{(k-2)!}\zeta(2k-3)(\frac{\beta_{R}}{4\pi})^{2(k-2)}\nonumber\\
&&+\frac{1}{64\pi^2}\int\limits_{0}^{\infty}
ids\, \ln(\frac{is}{8\pi^{2}G})\,\frac{\partial^3}{\partial(is)^3}
\Lambda(x,is)
\nonumber\\
&&-\frac{\alpha'}{64\pi G}R_{\mu\nu\tau\sigma}R^{\mu\nu\tau\sigma}
\exp[\,\frac{2\pi G}{3\beta_{R}^{2}}+\frac{G}{\pi}
\sum_{k=2}^{\infty}a_{k}(x)\frac{(2k-2)!}{(k-1)!}\zeta(2k-1)
(\frac{\beta_{R}}{4\pi})^{2(k-1)}\,]\; .
\label{eq:Ftot}
\end{eqnarray}
\narrowtext

\section{Finite temperature effective $\alpha'$ gravity action}

We identify $ -F[g]$ with the effective finite temperature
Lagrangian and  from Eq.\ (\ref{eq:DATr})
we find the modification  of the $\alpha'$ Einstein action by
thermal dilatons
\begin{eqnarray}
S_{eff}[g]=
&&\int d^{4}x\sqrt{-g}\,
(\frac{1}{16\pi G}R-F[g])\nonumber\\
=&&\int d^{4}x\frac{\sqrt{-g}}{16\pi\,G(\beta_{R})}(R-2\lambda(\beta_{R})+
\frac{\alpha'(\beta_{R})}{4}R_{\mu\nu\tau\sigma}R^{\mu\nu\tau\sigma})\; .
\label{eq:9}
\end{eqnarray}
We have defined the effective cosmological constant $\lambda(\beta_{R})$,
the gravitation constant $G(\beta_{R})$ and the effective string parameter
$\alpha'(\beta_{R})$
in high temperature limit
\begin{eqnarray}
&&\frac{1}{16\pi\;G(\beta_{R})}=
\frac{1}{16\pi G}+\frac{1}{6}\frac{1}{24\beta_{R}^{2}}\\
&&\frac{2\lambda(\beta_{R})}{16\pi\,G(\beta_{R})}
= -\frac{\pi^{2}}{90\beta_{R}^{4}}\\
&&\frac{\alpha'(\beta_{R})}{64\pi\,G(\beta_{R})}= \frac{\alpha'}{64\pi G}
\exp(\frac{2\pi G}{3\beta_{R}^{2}})\; .
\label{eq:Salp}
\end{eqnarray}
We will introduce the Planck mass $M_{P}=\frac{1}{\sqrt{G}}$ and
absolute temperature  $T_{R}=(\beta_{R})^{-1}$ in
Eq.\ (\ref{eq:Salp}) for the effective string parameter
$\alpha'(\beta_{R})$
\begin{equation}
\alpha'(\beta_{R})=
\alpha'\,\frac{\exp (\frac{2\pi G}
               {3\beta_{R}^{2}})}{1+\frac{\pi G}{9\beta_{R}^{2}}}=
        \alpha'\,\frac{\exp (\frac{2\pi}{3}\,\frac{T_{R}^{2}}{M_P^{2}})}
         {1+ \frac{\pi}{9}\,\frac{T_{R}^{2}}{M_{P}^{2}}}\; .
\label{eq:alpha}
\end{equation}

\section{Conclusion}

For the first time, we have derived the free energy density
(Eq.\ (\ref{eq:Ftot})) for the dilatons in external static gravitational field
using the RTFT methods to order $O(\alpha')$.

We have found related thermal modification of the $\alpha'$ Einstein
action. The temperature dependence of the effective string parameter
$\alpha'(\beta_{R})$ in high temperature region is given by
Eq.\ (\ref{eq:alpha}).
It is clear that the temperature $T_{R}$ has to be comparable with
the Planck mass $M_{P}$ to have some observable changes of the effective
parameter $\alpha'(\beta_{R})$.
The high temperature behavior  of  both the effective cosmological constant
$\lambda(\beta_{R})$ and the gravitational constant $G(\beta_{R})$ are similar
as in the case of the scalar pregeometry
which has been discussed in Ref. \cite{ThermGrav0}.

\acknowledgments
The author thanks J. Horsk\'{y}, M. Lenc and M. Pardy for discussions.
\newpage
\appendix

\section*{The real time Green's function}

In this appendix, we will derive  the RTFT Green's function for
the dilatons without the string corrections ($\alpha'=0$
in Eq.\ (\ref{eq:DilAc4})) in external static gravitational field.

We can rewrite the first term in the action (\ref{eq:DilAc4})
on the following form
\begin{eqnarray}
S[\phi]_{\alpha'=0}
=-\frac{1}{8\pi G} \int d^{4}x && \;
\sqrt{-g}\,\phi(x)(\Box+m^{2})\phi(x)\nonumber\\
=-\frac{1}{8\pi G}\int d^{4}x && \; [\,\phi(x)
({\eta^{\mu\nu}\partial_{\mu}\partial_{\nu}+m^{2}})\phi(x)\nonumber\\
&&+\phi(x)E(x^{i},\partial_{x})\phi(x)\,]\; ,
\label{eq:S}
\end{eqnarray}
where
\begin{eqnarray}
E(x^{i},\partial_{x})
=\sqrt{-g}\,(\Box+m^{2})
-(\eta^{\mu\nu}\partial_{\mu}\partial_{\nu}+m^{2})\; .
\label{eq:SE}
\end{eqnarray}
The second term in Eq.\ (\ref{eq:SE}) is calculated with help of the metric
$\eta^{\mu\nu}=diag(1,-1,-1,-1)$.
We have introduced helping mass term in the action (\ref{eq:DilAc4}).
We shall take limit $m \rightarrow 0$ in the  final equations.
It is important that the term $E(x^{i},\partial_{x})$
does not depend on the time variable $x^{0}$
because we consider static space-time.

Now we derive from the RTFT diagrammatic technique \cite{RTFT} RTFT-
causal Green's function $ \Delta[\beta](x,x')$ connected with the action
(\ref{eq:S})
\begin{eqnarray}
\frac{i\Delta[\beta](x,x')}{4\pi G}=
\sum_{n=0}^{\infty}i\Delta^{(0)}[\beta](E_{D}
i\Delta^{(0)}[\beta])^n ,
\label{eq:E}
\end{eqnarray}
where $E_{D}=-diag(E(x^{i},\partial_{x}),-E(x^{i},\partial_{x}))$ and
\begin{eqnarray}
i\Delta^{(0)}[\beta](k)
=U(\beta,k_{0}) \left( \matrix{ \frac{i}{k^2-m^2+i\epsilon} & 0 \cr
0 & \frac{-i}{k^2-m^2-i\epsilon} \cr } \right) U(\beta,k_{0})\;\; ,
\label{eq:Diag}
\end{eqnarray}
where
\begin{eqnarray}
U(\beta,k_{0})=
\left( \matrix{ \cosh \Theta_k & \sinh \Theta_k \cr
\sinh \Theta_k & \cosh \Theta_k \cr} \right) ,
\cosh^{2}(\Theta_{k})=\frac{1}{1-e^{-\beta|k_{0}|}}\; .
\label{eq:U}
\end{eqnarray}
Performing the sum (\ref{eq:E}) we obtain exact expression for
the Green's function
\widetext
\begin{eqnarray}
\frac{i\Delta[\beta](x,x')}{4\pi G}
=\int&& \frac{dk_{0}}{2\pi} \exp(ik_{0}(x^{0}-x'^{0}))\nonumber\\
&&\times U(\beta,k_{0})\; diag(\, i\Delta_{F}(k_{0})(x^{i},x^{'i}),
(i\Delta_{F}(k_{0})(x^{i},x^{'i}))^{*} \,)
 \;U(\beta,k_{0})\; .
\label{eq:DelDiag}
\end{eqnarray}
\narrowtext
We see  that because the external metric field has been considered
static it is possible to diagonalize the thermal dilaton Green's function.
The propagator $i\Delta_F$ is Feynman propagator in curved background
which can be expressed  in limit $x^{i}\rightarrow x'^{i}$ with using
relevant DeWitt coefficients
$a_{k}(x)$
\cite{BirDev,BunPar}
\widetext
\begin{eqnarray}
i\Delta_F(x,x')=&&
\int \frac{dk_{0}}{2\pi}e^{ik_{0}(x^{0}-x'^{0})}
i\Delta[\beta]_F(k_{0})(x^{i},x^{'i})\nonumber\\
=&&\int \frac{d^{4}k}{(2\pi)^4} e^{ik_{0}\sqrt{g_{00}}(x^{0}-x'^{0})}
\sum_{n=0}^{\infty}\,a_{n}(x)\,
(-\frac{\partial}{\partial m^{2}})^{n}\,\frac{i}{k^2-m^2+i\epsilon}\; .
\label{eq:Da(xx)}
\end{eqnarray}
Now we can express the coincidence limit for $\Delta[\beta]_{11}(x,x)$
using Eqs.\ (\ref{eq:DelDiag}),(\ref{eq:Da(xx)})
\begin{eqnarray}
&&\frac{i\Delta[\beta]_{11}(x,x)}{4\pi G}\nonumber\\
&&=\int \frac{d^{4}k}{(2\pi)^4}\sum_{n=0}^{\infty}\,a_{n}(x)\,
(-\frac{\partial}{\partial m^{2}})^{n}
[\frac{i}{k^2-m^2+i\epsilon}
+\frac{2\pi}{\exp(\beta\sqrt{g_{00}}|k_{0}|)-1}\,
\delta(k^{2}-m^{2})]\nonumber\\
&&=\frac{1}{16\pi^{2}}\int\limits_{0}^{\infty} \frac{ds}{s^{2}}\exp(-sm^{2})
\Lambda(x,s)\sum_{n=-\infty}^{\infty}\exp(\frac{-{\beta_{R}}^2 n^{2}}{4s})\; ,
\label{eq:Delta}
\end{eqnarray}
\narrowtext
where
\begin{equation}
\Lambda(x,s)\equiv \sum_{k=0}^{\infty}a_{k}(x)s^{k}\\
\label{eq:DeltaL}
\end{equation}
and
\begin{eqnarray}
a_{0}(x)=&&1 \; , \; a_{1}(x)=\frac{1}{6}R\; ,\nonumber\\
a_{2}(x)=&&
\frac{1}{72}R^{2}
+\frac{1}{180}(-R_{\mu\nu}R^{\mu\nu}+
R_{\mu\nu\tau\sigma}R^{\mu\nu\tau\sigma})
-\frac{1}{30}\Box R\; .
\label{eq:a(xx)}
\end{eqnarray}
We  have written out explicitly the first three DeWitt coefficients
in Eq.\ (\ref{eq:a(xx)}). The inverse temperature $\beta_{R}$ is the
local Lorenz-rest-frame inverse temperature given by Eq.\ (\ref{eq:betaR}).

\end{document}